\documentclass[runningheads]{cl2emult}

\usepackage{makeidx}  
\usepackage{graphicx} 
\usepackage{subeqnar} 
\usepackage{multicol} 
\usepackage{cropmark} 
\usepackage{eso}      
\makeindex            

\newcommand{\bld}[1]{\mbox{\boldmath $#1$}}
\newcommand{\et}[1]{\, \mathrm{e}^{\mbox{\footnotesize $#1$}}\, }
\newcommand{\half}{{\textstyle \frac{1}{2}}}

\begin{document}
\title*{Regularization and Inverse Problems}
\toctitle{Regularization and Inverse Problems}
%
%
\titlerunning{Regularization and Inverse Problems}
%
\author{Anthony Lasenby \and Bel\'en Barreiro \and Michael Hobson}
\authorrunning{Lasenby et al.}
%
%
\institute{Astrophysics Group, Cavendish Laboratory, Madingley Road,\\
Cambridge, CB3 0HE, U.K.}

\maketitle              

\begin{abstract}
An overview is given of Bayesian inversion and
regularization procedures. In particular,
the conceptual basis of the maximum entropy
method (MEM) is discussed, and extensions to
positive/negative and complex data are highlighted. Other deconvolution
methods are also discussed within the Bayesian context, 
focusing mainly on the comparison of
Wiener filtering, Massive Inference and the Pixon method, 
using examples from both astronomical and 
non-astronomical applications.
\end{abstract}

\section{Introduction}

In the next few years there will exist all-sky datasets from two new
satellite missions for the Cosmic Microwave Background (the MAP and
Planck missions), along with very large datasets from optical
surveys such as 2dF and Sloan. The combined effect of these new
data on quantitative cosmology will be enormous, but at the same time
pose great problems in terms of the scale of data analysis effort
required. 

As an example, the Planck Surveyor
satellite, due for launch in 2007,
combines both HEMT and bolometer technology in 10 frequency
channels covering the range 30~GHz to 850~GHz, with a highest
angular resolution of 5 arcmin. An artist's impression of this satellite 
is shown in Figure~\ref{fig:Planck-pic},
\begin{figure}[t]
\caption{Artist's impression of the Planck Satellite}
\label{fig:Planck-pic}
\end{figure}
and the experimental parameters of the Planck mission are summarized
in Table~\ref{tab:planck-params}.
\begin{table}
\centering
\caption{Approximate experimental 
parameters of the Planck satellite. HFI 
refers to the high frequency part
of the instrument, and LFI is the low frequency instrument. 
The $\Delta T/T$ 
sensitivity
is per beam area in one year (thermodynamic temperature)} 
\renewcommand{\arraystretch}{1.4}
\setlength\tabcolsep{5pt}
\begin{tabular}{|l|c|c|c|c|c|c|c|c|c|c|}
\hline
& \multicolumn{4}{|c|}{LFI (HEMT)} & \multicolumn{6}{|c|}{HFI (Bolometers)} \\ \hline
$\nu$ GHz & 30 & 44 & 70 & 100 & 100 & 143 & 217 & 353 & 545 & 857 \\ 
No. of detectors & 4 & 6 & 12 & 34 & 4 & 12 & 12 & 6 & 8 & 6 \\ 
$\theta_{\rm FWHM}$ & $33'$ & $23'$ & $14'$ & $10'$ & $10.7'$ & $8'$ & $5.5'$ & $5'$ & $5'$ & $5'$ \\ 
$\Delta T/T \times 10^{-6}$ & 1.6 & 2.4 & 3.6 & 4.3 & 1.7 & 2.0 & 4.3 & 14.4 & 147 & 6670 \\ 
Polarization & yes & yes & yes & yes & no & yes & yes & yes & no &
no \\ 
\hline
\end{tabular}
\label{tab:planck-params}
\end{table}
The mission is designed to give high sensitivity to CMB structures, 
together with
sufficient frequency coverage to enable accurate separation of the non-CMB 
physical components. These will typically be Galactic dust, synchrotron and 
free-free emission,  together with extragalatic radio and sub-mm/FIR sources.
Also present will be 
the effects of Sunyaev-Zeldovich distortion of the CMB as it passes through the
hot intracluster gas of clusters of galaxies. This separation of components
must be performed 
using data from approximately 100 detectors in total, spanning ten
frequencies, and with the sky map at each frequency
containing on the order of $10^7$ pixels. 
These figures give some idea of the scale of the problem, for just this
mission alone, and suggest why the idea of `mining the sky' is
appropriate.

The task of analysing modern large datasets is undeniably challenging
in terms of the amount of data to be processed. In the pursuit of
`precision cosmology', however, we are faced with the additional
requirement that the data must be analysed in a statistically rigorous
way. In CMB observations, for example, one is interested 
in the statistical properties
of CMB anisotropies, most commonly summarised by their power spectrum
$C_\ell$, from which it is possible to derive estimates and confidence
limits on fundamental cosmological parameters such as the matter
density of the Universe or the value of the cosmological constant.
Similar statistical measures are central to the analysis of optical
surveys. Thus, in modern cosmology, one is faced with the  
dual problem of analysing large datasets while retaining statistical
rigour. In the present paper, we discuss both aspects, particularly in
the context of how an efficient choice of `basis functions' can lead
to both an improved analysis and large speed-up factors.

It is now generally accepted that the correct way to draw inferences
from any set of data is to apply Bayes' theorem in a consistent and 
logical manner. This provides a general framework in which 
the analysis of CMB and optical survey data can be performed.
Let us consider the generic problem at hand.
In order to recover an underlying signal $\bld{s}$ from some 
measured data $\bld{d}$, we commonly need to solve an inverse problem 
such as
\begin{equation}
\bld{d} = R \bld{s} + \bld{\epsilon},
\end{equation}
where $R$ represents the response matrix of the experiment and 
$\bld{\epsilon}$ is the instrumental noise vector. For simplicity, 
we are assuming here that the inversion problem is linear, although
this is not strictly necessary. In any case,
owing to the presence of noise, the properties of which are
only known statistically (sometimes even this is not true), the
inversion problem is degenerate.
Even in absence of noise, a direct inversion would, in general, not be
possible, since the response matrix $R$ is normally not
invertible. For instance, $R$ may be a blurring (beam)
function, which strongly suppresses higher spatial frequencies, 
or it may represent a
beam-differencing experiment where some spatial frequencies are
actually set to zero. Thus, it is clear that some kind of statistical
technique is needed in order to regularise the inversion.
This naturally leads us to a Bayesian approach. This is one of the
most powerful current techniques of image reconstruction.

In the present paper, we discuss different deconvolution methods
within the Bayesian framework, showing that different techniques are
actually obtained by different choices of priors and/or basis
functions. The outline of the paper is as follows. \S\ref{bayes} gives an
introduction to Bayes' theorem and derives the Wiener filter in this
context. \S\ref{mem} describes the Maximum Entropy Method (MEM), including
extensions to positive/negative and complex data, and
discusses some applications. The
Pixon Method is introduced in \S\ref{pm}. \S\ref{wmem} discusses 
multiscale and wavelet MEM.
The Massive Inference technique is introduced in
\S\ref{minf}. Finally, conclusions are given in
\S\ref{conc}.

\section{Mining the Sky with Bayes' Theorem}
\label{bayes}

Let us recall the original problem
\begin{equation}
\bld{d} = R \bld{s} + \bld{\epsilon},
\end{equation}
For simplicity we assume
$\langle \bld{s} \rangle = \bld{0} = \langle \bld{\epsilon} \rangle$

To obtain the `best' sky reconstruction we chose to maximise the
probability $Pr(\bld{s}|\bld{d})$ using {\it Bayes' theorem}

\begin{equation}
Pr(\bld{s} | \bld{d}) = \frac{1}{Pr(\bld{d})} \, Pr(\bld{d} | \bld{s}) \, Pr (\bld{s}),
\end{equation}
where $Pr(\bld{s} | \bld{d})$ is the posterior probability of an underlying
signal (or true sky) $\bld{s}$ given 
some data $\bld{d}$, $Pr(\bld{d} | \bld{s})$ is the likelihood
funtion and $P(\bld{s})$ is the prior probability. 
At the first level of Bayesian inference $Pr(\bld{d})$, the evidence, 
is merely a normalisation, which implies we wish to maximise
\begin{equation}
Pr(\bld{s} | \bld{d}) \propto Pr(\bld{d} | \bld{s}) \, Pr (\bld{s})
\end{equation}
 
For convenience we consider the case of Gaussian noise, although this
is not necessary (for instance there exist many applications to Poisson
noise). For Gaussian noise, the likelihood is simply
\begin{equation}
Pr(\bld{d} | \bld{s}) \propto \et{-\half \bld{\epsilon}^T N^{-1}
\bld{\epsilon}}  
= \et{-\half(\bld{d} - R \bld{s})^T N^{-1} (\bld{d} - R \bld{s})}
\end{equation}
where $N = \langle \bld{\epsilon} \bld{\epsilon}^T \rangle$ is the
noise covariance 
matrix. This is usually written as $Pr(\bld{d} | \bld{s}) \propto 
\exp(-\half\chi^2)$. Now we have to decide on the assignment of the
prior, $Pr(\bld{s})$. As a first approach, we assume that $\bld{s}$ 
(which, for a CMB experiment, for example, would include 
CMB anisotropies, the Sunyaev-Zeldovich effect, Galactic emission,
etc.) is a Gaussian random variable, described by a known
covariance matrix $C = \langle \bld{s} \bld{s}^T \rangle$ (including all
cross-correlations) so that
\begin{equation}
Pr(\bld{s}) \propto \et{-\half \bld{s}^T C^{-1} \bld{s}}
\end{equation}
In this case the posterior probability is
\begin{equation}
Pr(\bld{s} | \bld{d}) \propto Pr(\bld{d} | \bld{s}) \, Pr (\bld{s})
\propto \et{-\half (\chi^2+\bld{s}^T C^{-1} \bld{s})}
\end{equation}
which one must maximise with respect to $\bld{s}$ to obtain the reconstruction.
This is equivalent to minimising $F = \half (\chi^2+\bld{s}^T C^{-1} \bld{s})$.
In fact, we can do better than this. By completing the square in \bld{s}
(e.g.~\cite{z95}), we can recover the whole posterior distribution:

\begin{equation}
Pr(\bld{s} | \bld{d}) \propto \et{-\half (\bld{s}-\hat{\bld{s}})^T E^{-1} (\bld{s} - \hat{\bld{s}})}
\end{equation}
where the sky reconstruction $\hat{\bld{s}}$ is given by
\begin{eqnarray}
\hat{\bld{s}} &=& W \bld{d}, \nonumber \\
W &=& (C^{-1} + R^T N^{-1} R)^{-1} R^T N^{-1},
\end{eqnarray}
where $W$ is in fact the Wiener matrix and
\begin{equation}
E = (C^{-1} + R^T N^{-1} R)^{-1}
\end{equation}
is the reconstruction error matrix $E = \langle
(\bld{s}-\hat{\bld{s}})(\bld{s}-\hat{\bld{s}})^T\rangle$.
Thus we have recovered the optimal linear method, which is
usually derived by minimising residual variances.

We recall that in general the response matrix $R$ will not be
invertible. 
However, it is remarkable that 
the estimation of the sky $\hat{\bld{s}} = W \bld{d}$ can still
be computed no matter how singular $R$ is, since it only needs
$R^T$ to be evaluated. This is an example of regularization.
Notice how if the $C^{-1}$ were not present in $W$ we would just have
$W=R^{-1}$. We say that we have regularized the inverse.

The above solution is `easy' to calculate and has known reconstruction errors. 
It is, however, by no means the best solution in real problems. For
instance, consider the standard `Lena' IEEE test image in Fig.~\ref{lena}. The
original image (top left panel) is smoothed with a Gaussian blurring
function with a FWHM of 6 pixels followed by the addition of noise
(top right panel). The Wiener filter reconstructed image is given in
the bottom right panel. Although some improvement is achieved,
spurious structure (`ringing') appears at small scales.
For comparison, a result generated using a pixon method (see~\S\ref{pm})
is also shown.
\begin{figure}
\label{lena}
\caption{Comparison of the performance of a pixon method to the Wiener
filter for the `Lena' test image. The original image has been
blurred with a Gaussian blurring function with a FWHM of 6 pixels followed by
addition of noise}
\end{figure}

\section{The Maximum Entropy Method}
\label{mem}

The main shortcoming of the Wiener filter is that relies on the assumption of
Gaussianity and the a priori knowledge of the covariance
matrix. Real data, however, is rarely so simple, and we must therefore
consider alternative priors. A possible choice is the entropy prior 
(Maximum Entropy Method, MEM).

Usually MEM is applied to positive, additive distibutions (PADS). 
Let $\bld{h}$ be the (true) pixel vector we are trying to estimate. In
this case very general considerations of subset independence,
coordinate invariance and system independence 
lead uniquely to the prior $Pr(\bld{h}) \propto \et{\alpha S}$ where
the `entropy' $S$ (\cite{s89}) of the image is given by
\begin{equation}
S(\bld{h},\bld{m}) = \sum_i \left(h_i - m_i -h_i \ln\left(\frac{h_i}{m_i}\right)\right)
\end{equation}
where $\bld{m}$ is the measure on an image space (the model) to which
the image $\bld{h}$ defaults in the absence of data (it can be shown
that the global maximum of $S$ occurs at $\bld{h}=\bld{m}$).
In fact, it has been shown recently (\cite{anton}) that,
if there exist linear constraints on the signal
(e.g. like $\bld{d} = R \bld{s} + \bld{\epsilon}$ in our case),
the form of the entropic prior is determined uniquely by simply requiring
consistency with the sum and
product rules of probability.

`Subset independence' 
implies, however, that no {\em a priori\/} correlations between the
pixels of $\bld{h}$ should be present. 
So, is it possible to include known covariance
structure, as in the Wiener method?. The answer is yes!. 
Given a sky $\bld{s}$ with $C=\langle \bld{s} \bld{s}^T 
\rangle$, we form the Cholesky decomposition
\begin{equation}
C = L L^T 
\end{equation}
where $L$ is an upper triangular matrix,
and define a hidden, uncorrelated i.i.d. (independent, identically
distributed) unit variance hidden field $\bld{h}$ related to $\bld{s}$ by
\begin{equation}
\bld{s}=L\bld{h}
\end{equation}
It is straightforward to show that, with this construction,
$\langle \bld{s}\bld{s}^T\rangle =C$.
Thus the derivation of $S(\bld{h},\bld{m})$ applies to this hidden
variable and we need to maximise
\begin{eqnarray}
Pr(\bld{h} | \bld{d})& \propto &
\et{-\half \chi^2(\bld{h})+\alpha S(\bld{h},\bld{m})} \nonumber
\\
{\rm where}~~~ \chi^2(\bld{h})&=&(\bld{d}-RL\bld{h})^T N^{-1}
(\bld{d}-RL\bld{h})
\end{eqnarray}
The vector $\hat{\bld{h}}$ that maximises this expression is the
MEM reconstruction.
Note that $\alpha$ is a regularising parameter of the relative weight
of the data and the prior. Large $\alpha$ favours large entropy (i.e.,
$\hat{\bld{h}}$ close to $\bld{m}$) at expense of the data, 
whereas small $\alpha$ gives more weight to the data.
The parameter 
$\alpha$ can be estimated itself via Bayesian methods (\cite{s89}).
Crudely, the value of $\alpha$ is such that
\begin{equation}
\chi^2(\hat{\bld{h}}) \approx N,
\end{equation}
where $N$ is the number of good degrees of freedom in the data.
Note that for $\alpha=0$, the method reduces to maximum likelihood.
For $\alpha=2$ and small $h_i$ (in fact for $h_i 
\mathrel{\hbox to 0pt{\lower 3.5pt\hbox{$\mathchar"218$}\hss}\raise 1.5pt\hbox{$\mathchar"13C$}} 3m, \forall_i$)
it can be shown that the method is Wiener filter again (\cite{h98}).
This means that the Wiener filter is simply a quadratic 
approximation of MEM with $\alpha=2$.

Another important issue is how to calculate the errors on the
reconstruction. This is performed by making a 
Gaussian approximation to the posterior probability
distribution $Pr(\bld{h} | \bld{d})$ at its peak $\hat{\bld{h}}$. 
Moreover, by sampling from this distribution
within, say, the $1\sigma$ surface, one can generate sample
reconstructions all compatible with the data, which can be very
informative.

\begin{figure}
\label{gravl}
\caption{The shear field in the direction of the galaxy cluster
MS1054, determined from HST observations, and the corresponding
MEM reconstruction of the projected mass density in the cluster
(courtesy of Phil Marshall)}
\includegraphics[width=5.5cm,angle=-90]{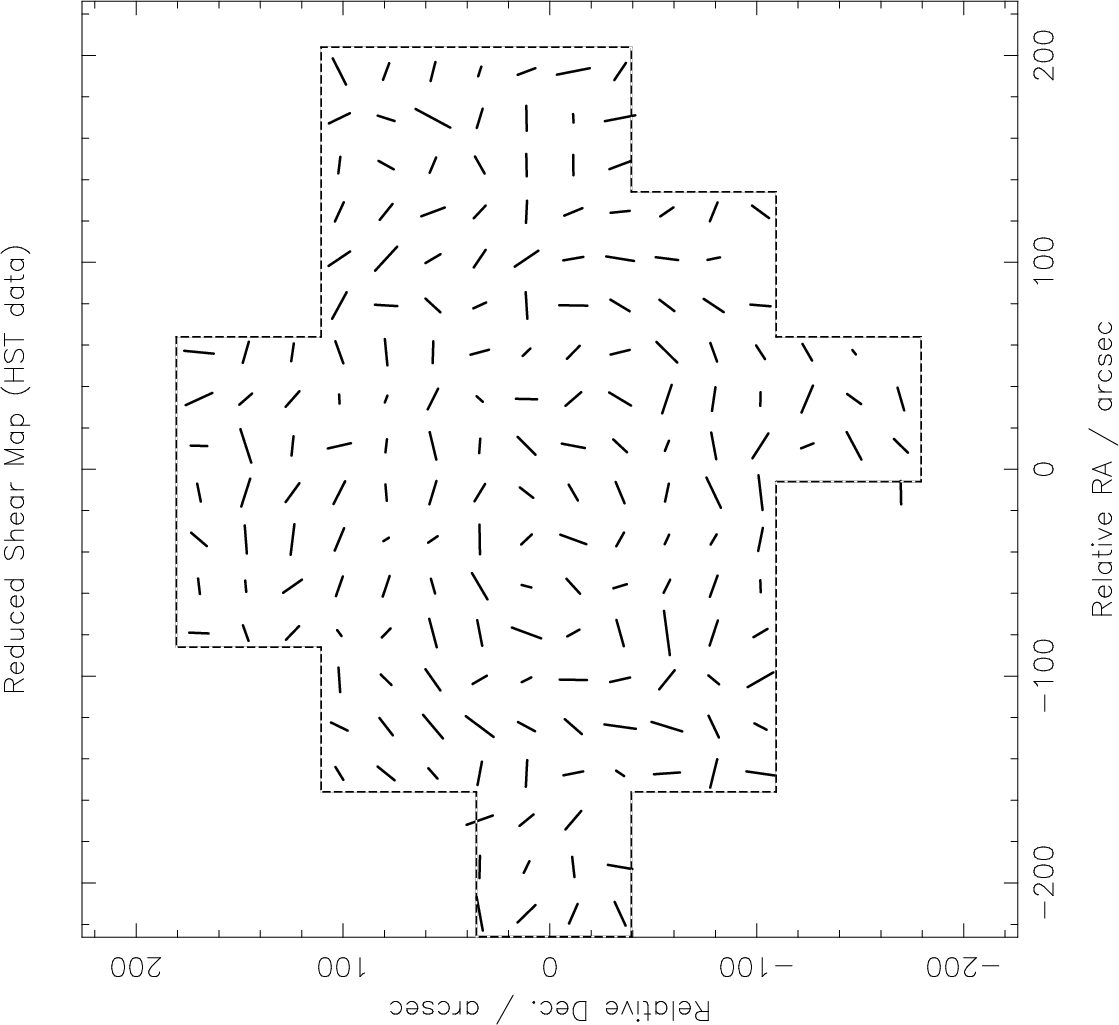}
\qquad
\includegraphics[width=5.5cm,angle=-90]{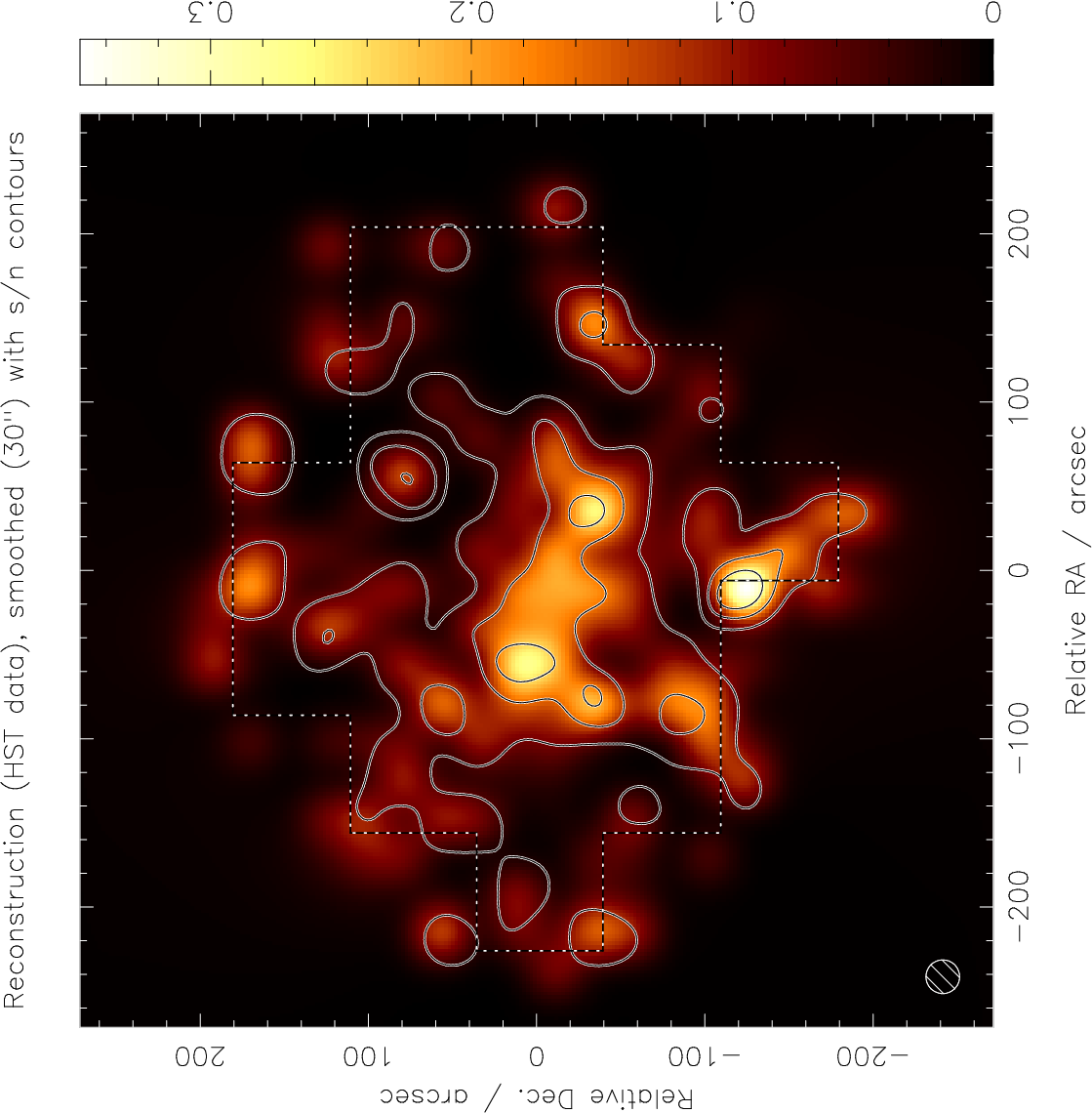}
\end{figure}

An interesting application of MEM to astronomical data is the
recovery of the projected mass density of a galaxy cluster from
observations of its gravitational lensing effects on background
galaxies (\cite{b98}). This technique is particularly interesting since 
it directly maps the dark
matter halos in clusters. Moreover, together with the projected mass
distribution, an estimation of errors is also obtained. Figure~\ref{gravl}
shows the projected mass density of the cluster 
MS1054 reconstructed from shear data obtained 
by \cite{hoek} using the Hubble Space Telescope.

A further extension of MEM is necessary in order to apply the algorithm
to positive and negative data (such as CMB) and also to complex data
(e.g. Fourier transforms). Indeed, it is possible to generalise MEM to both
of these kinds of data.
For a positive and negative image, we just need to write $\bld{h}$ as the
difference between two positive images
\begin{equation}
\bld{h}=\bld{u}-\bld{v}
\end{equation}
Applying continuity constraints, we then obtain the entropic prior for
positive/negative images as
\begin{eqnarray}
S(\bld{h},\bld{m})&=&\sum_i \psi_i -2m_i- h_i \ln\left( \frac{\psi_i+h_i}{2
m_i}\right) \nonumber \\
{\rm with~~} \psi_i &=& \sqrt{h_i^2+4m_i^2}.
\end{eqnarray}
The posterior probability is given, as before, by
$\exp(-\half \chi^2 +\alpha S)$, 
but now using this generalised definition of entropy.
This result can be derived directly from
counting  arguments (`monkeys throwing balls') \cite{hl98}.
Regarding complex images, we can just treat real and imaginary parts
separately:
\begin{equation}
\label{scompleja}
S(\bld{h},\bld{m})= S(\Re(\bld{h}),\Re(\bld{m})) +
S(\Im(\bld{h}),\Im(\bld{m}))
\end{equation}
where $\Re$ and $\Im$ denote the real and imaginary parts of each
vector.
This generalisation to positive/negative and complex images 
is actually a key point, since MEM now can be applied to
the Fourier Transform (or Spherical Transform for the all-sky case) 
of the original maps or images.
But in Fourier space, modes at different $\bld{k}$ (or $l,m$ on a sphere) can
generally be treated independently, therefore we can apply MEM
separately at each mode. This means that we have $N_{pix}$
minimisations with respect to one or a few variables, instead of a
single minimisation with respect to $N_{pix}$ variables. This leads to a
huge speed-up in the algorithm, which is crucial for large data sets.

We call this FastMEM or FourierMEM. This method has been successfully
applied to reconstructing the different components of the microwave sky
from simulated Planck data of small patches of the sky (\cite{h98}).
An application of FastMEM to Planck data is also given
in this volume (\cite{b01}).
Moreover, an extension of the algorithm to deal with all-sky data,
which works in spherical harmonic space, is
currently being tested (\cite{vlad}).
\begin{figure}
\caption{Results of MEM as applied to Planck simulated data on the
whole sky. From top to bottom the maps correspond to input CMB, input
dust and residuals for the MEM reconstructed CMB (from \cite{vlad})}
\label{sphere}
\end{figure}
Fig.~\ref{sphere} shows the performance of this technique
for simulated all-sky Planck data. The input CMB and Galactic dust maps are
shown, but in addition the simulations also contain Galactic
synchrotron and free-free emission as well as thermal and kinetic
Sunyaev-Zeldovich effect from clusters of galaxies. The bottom panel
shows the residuals in the MEM reconstruction of the CMB. 
It is striking that, even when no cut of the Galactic plane has been
attempted, no obvious emission from the Galaxy has contaminated the
reconstructed CMB, except for a few pixels in the centre of the map.

\begin{figure}
\caption{Blurred image of a pollen grain and FastMEM reconstruction}
\label{pollen}
\includegraphics[width=5.3cm,angle=-90]{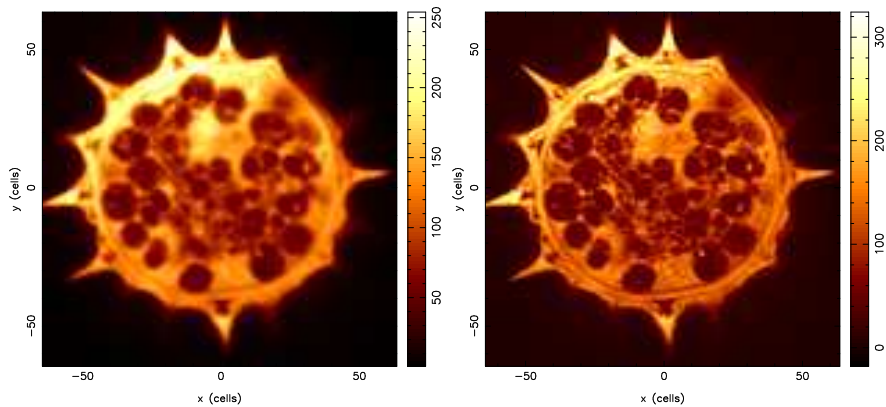}
\end{figure}

FastMEM has also performed very well in non-astronomical
data. The left panel of Fig.~\ref{pollen} shows the blurred image 
(due to the instrument response) 
of a pollen grain obtained by combining 20 images taken at different 
depths with a confocal microscope. The reconstructed
image achieved by FastMEM is shown in the right panel. The amount of
detail recovered with respect to the original image is very
noticiable. Besides, FastMEM takes around 45 seconds to perform such
a reconstruction versus 50 minutes needed by real spece MEM.

\section{The Pixon Method}
\label{pm}

A recent addition to the stable of image reconstruction algorithms is
the pixon method (\cite{p93}). 
The basic idea behind this technique is to minimise the number of
degrees of freedom used to describe an image while still
maintaining an acceptable fit to the data. 
This is achieved by, instead of working in the pixel basis, describing 
the image using `pixons',
which are essentially flexible pixels able to
change shape and size. For example, in
the pixon approach, only a few large pixons are needed to describe the
background or parts of the image with a low signal to noise ratio,
whereas a larger number of smaller pixons are used where 
the signal has more detail.

The pixon idea can in fact be phrased within in the framework of
Bayes' theorem via the introduction of a prior. 
Let there be $N$ counts (e.g. photons) in total that must be
assigned to $n$ cells (or pixons) and assume $N_i$ counts end up in
pixon $i$ , so $\sum_i N_i =N$. Thus, we need to choose $n$ and $N_i$
and also the position of the pixons in the least informative way.
The total number of possibilities is given by $n^N$. The total number
having a given $N_1$ in pixon $1$, $N_2$ in pixon $2$, etc. is
$\frac{N!}{\Pi_iN_i!}$. So, the probability of a given arrangement is
\begin{equation}
Pr({\rm arrangement})=\frac{N!}{n^N \Pi_i N_i!}
\end{equation}
Note that this probability favours arrangements with a small number of
pixons containing a large number of counts instead of having a large
number of cells with only few counts. Indeed, it is maximised by
$n=1$ and $N_i=N$. 
So, we can use this probability as a prior combined with the
likelihood term to obtain the posterior probability.
Moreover, using Stirling's approximation, we can write this as
\begin{equation}
Pr({\rm arrangement}) \approx \frac{1}{n^N}\exp \left(-\sum_i
\frac{N_i}{N} \ln \frac{N_i}{N}\right)
\end{equation}
which is similar to an entropy prior. Thus, the pixon method can be
seen as a MEM that allows `pixel sizes' (and shapes) to vary as well.

Note that we have described the `pure form' of the pixon method, but
so far the commercial code has had to include a large number of
modifications relative to this in order to get an algorithm that works
properly and rapidly enough.
An independent implementation of the Pixon method for cluster
detection is given by~\cite{eke} in this volume. Indeed, the notion of
distinct `hard-edged' pixons of different shapes and sizes is 
unhelpful in the reconstruction of general images, and current pixon
algorithms tend to favour a `fuzzy-pixon' approach, which is equivalent
simply to the assumption of an instrinsic correlation length for
the structure in the image, which can vary across the image.
Thus, the reconstructed image $I$ is written as the local convolution of a
pseudo-image $I_{\rm pseudo}$ with a pixon shape function $K$, whose
width varies over the image
\begin{equation}
I(\bld{x}_i)=\int_{V_{\bld{y}}}
K\left(\frac{\bld{y}-\bld{x}_i}{\delta_i}\right) 
I_{\rm pseudo}(\bld{y}) dV_{\bld{y}},
\end{equation}
where $\bld{x}_i$ is the location of pixel $i$ and $\delta_i$ is the 
pixon size at pixel  
$i$. The pixon shape can be arbitrary (which can be a strength or
a weakness of the method in different circumstances).
A common choice is a truncated inverted paraboloid (\cite{p94}), 
which leads to (except for a normalisation)
\begin{equation}
K= \left\{ \begin{array}{ll}
1-\frac{|\bld{y}-\bld{x}_i|^2}{\delta_i^2}, & |\bld{y}-\bld{x}_i| \le \delta_i \\
0, & |\bld{y}-\bld{x}_i| > \delta_i \\
\end{array} \right.
\end{equation}

The basic algorithm is very simple. Firstly, some initial choice is
made for the pixon width $\delta_i$ in each pixel. Often this can be
simply $\delta_i=1$ for all $i$, but this can lead to `freezing-in'
of unwanted small-scale structure, so in some cases the $\delta_i$ are
chosen to be somewhat larger. In any case, given the initial choice of
the $\delta_i$ the maximum-likelihood solution for the pseudo-image
$I$ is obtained in a standard manner. Then, keeping this pseudo-image
fixed, the pixon widths $\delta_i$ are varied until a set is found
where each $\delta_i$ has the  largest possible value that is
still consistent with the data in a least-squares sense. This whole
two-step process is then repeated until convergence is achieved.

\begin{figure}
\caption{Comparison of the performance of the Pixon method to
traditional MEM. Top row (from
left to right): original image, blurred and noisy image, Pixon
reconstructed image, MEM reconstruction. Centre row: surface plots for
the images in top raw. Bottom row: blurring function, additive noise,
residuals for Pixon Method, residuals for MEM. (Images from 
{\tt http://casswww.ucsd.edu/personal/puetter/pixonpage.html)}}
\label{pixon}
\end{figure}
Fig.~\ref{lena} showed a comparison between the pixon method and
the Wiener filter. We see that the pixon method clearly outperforms Wiener
filter. Another example is given in Fig.~\ref{pixon}.
In this case, a synthetic image composed of a sharp peak
and a valley is used to compare the pixon method and MEM. 
The image has been blurred and Gaussian noise added.
We see that the recovery of the peak and valley are similar for
both Pixon and MEM, but the low level noise present in the background
of the MEM reconstruction has been successfully removed in the Pixon
image. In addition, the residuals for the Pixon method seem
compatible with random noise whereas MEM produces residuals correlated
with the signal.

\section{Multiscale and wavelet MEM}
\label{wmem}

Although the reconstructions in Figure~\ref{pixon} show
the pixon method to be very effective, the comparison
is not strictly reasonable, since it employs the
traditional MEM technique, which is rarely still used in this
simple form. 

It has long been realised that the key to effective
image reconstruction is to reduce the number of degrees of
freedom one is trying to constrain. The simplest way of
achieving this goal is via the assumptions of an intrinsic
correlation length that does not vary across the image 
(\cite{skg}); this was
discussed briefly in \S\ref{mem}. Basically, one
hypothesises the existence of a hidden space $\bld{h}$ that is
linearly-related to the signal space $\bld{s}$ by an intrinsic
correlation function $L$, such that
\begin{equation}
\bld{s} = L\bld{h}.
\end{equation}
One then performs the MEM reconstruction in terms of $\bld{h}$ (which
is a priori uncorrelated). The
corresponding signal reconstruction $\bld{s}$ will thus have an intrinsic
correlation length determined by $L$ and hence fewer independent
degrees of freedom. This clearly
corresponds to the pixon method with all the pixon widths 
$\delta_i$ being equal.

\begin{figure}
\begin{center}
\caption{The multiscale MEM reconstruction of the projected mass 
density in the cluster MS1054 (courtesy of Phil Marshall)}
\label{gravl2}
\includegraphics[width=6cm,angle=-90]{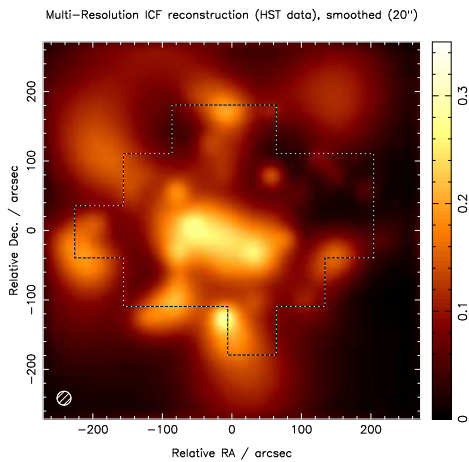}
\end{center}
\end{figure}

However, this simple method can be greatly enhanced by choosing 
$L$ in a more innovate way. The most obvious extension is to allow
for the existence of multiple hidden fields, each related to signal
space by convolutions of different widths. This then reproduces the
ability to have varying effective correlation lengths across the
image, but in a such way that the the correlation length at each point
in the image is determined
via a proper entropic regularisation of the hidden fields, and not
by an arbitrary least-squares criterion as is done in the pixon method.
Application of this `multiscale MEM' technique to numerous types of 
images has shown it to be very successful. An interesting astronomical
example is again provided by gravitational
lensing. Figure~\ref{gravl2}
shows the reconstruction of the projected mass density in the cluster 
MS1054 from the shear data shown in Figure~\ref{gravl}. 
In this case,
however,
the reconstruction has been performed using a 4-level multiscale MEM
algorithm. By comparing with the traditional MEM reconstruction in
Figure~\ref{gravl} one sees that the small scale rippling has
disappeared, and indeed the calculated evidence for the 4-scale 
reconstruction is much higher. By way of illustration,
in Figure~\ref{gravl3} we also plot
the corresponding hidden fields that constitute the reconstruction.
\begin{figure}
\caption{The hidden fields that constitute the multiscale MEM
reconstruction in Figure~\ref{gravl2} of MS1054 
(courtesy of Charlie McLachlan)}
\label{gravl3}
\end{figure}

One can see from figure~\ref{gravl3} that the mutliscale MEM approach is
equivalent to providing a set of (redundant, non-orthogonal) 
basis functions for the image that
are simply the different intrinsic correlation functions. 
The MEM is simply obtaining a properly regularised optimal solution
for the values of the coefficients of each basis function required
to reconstruct the image.
Once viewed
in this way, one may wonder if there exist more efficient sets of
basis functions one could use to describe the image. Clearly, the
number of degrees of freedom is simply equal to the number of basis
functions required, and so one wishes to find a basis in which 
general images can be described with relatively few basis
functions. The obvious choice is wavelets. These functions are
constructed so that they are well-localised in both position and
frequency space, and have proven to be very effective in representing
an image with few basis functions (their extensive use in image
compression is also obviously a result of this property). Indeed,
by using a wavelet transform kernel to relate the spaces $\bld{h}$
and $\bld{s}$, the reconstruction quality can be improved still further.

\section{Massive Inference}
\label{minf}

Massive Inference (\cite{sk98},\cite{sk99}) can be seen as an even
more extreme choice of basis functions. 
In this method, we throw away the underlying pixelisation
grid and instead represent the object as a variable number of `atoms'
or `point masses'.
Each of these is described by a 
position $\bld{x}_j$ and a flux $z_j$. To simulate a 
continuum, $\bld{x}_j$ runs over $2^{32}$
positions. We need to assign a prior probablity to $\bld{x}$ and $z$
as well as to the number of atoms $N$.

Each of the $N$ locations is assigned a uniform prior, i.e.,
$Pr(\bld{x}_i)= \mbox{constant}$. 
The number of atoms can be assigned a Poisson
distribution with a given mean $\alpha$:
\begin{equation}
Pr(N)=\frac{\et {-\alpha}\alpha^N}{N!}
\end{equation}
Finally, each of the $N$ amplitudes is assigned an exponential prior
(with parameter $q$):
\begin{equation}
Pr(z_i)=\frac{\et {-z_i/q}}{q}
\end{equation}
The program (using Markov Chain Monte Carlo sampling and simulated annealing)
then samples the posterior probability (which also includes the
likelihood term) treating $\alpha$ and $q$ as hyperparameters.

\begin{figure}
\begin{center}
\caption{The decay of luminescence in a simulated flash photolysis experiment}
\label{flash1}
\includegraphics[width=4cm,angle=-90]{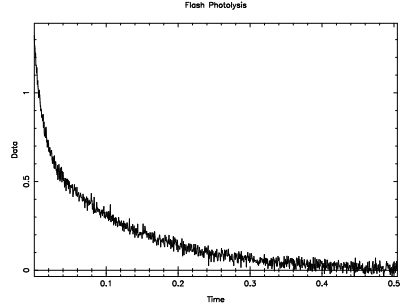}
\end{center}
\end{figure}

\begin{figure}
\begin{center}
\caption{The decaying exponential components present in
Figure~\ref{flash1} as determined by MassInf (left) and MEM (right) 
respectively}
\label{flash2}
\includegraphics[width=4cm,angle=-90]{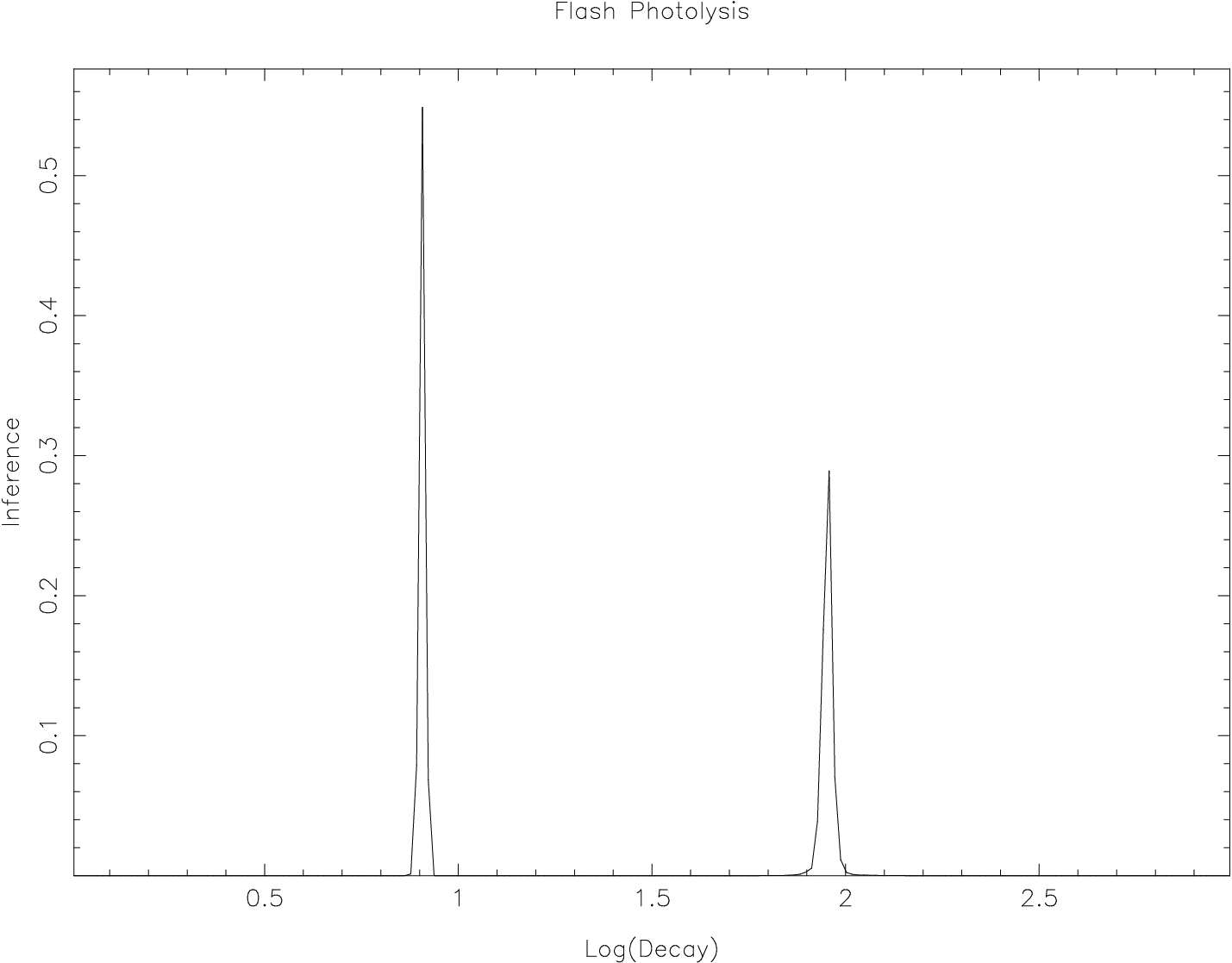}
\qquad\includegraphics[width=4cm,angle=-90]{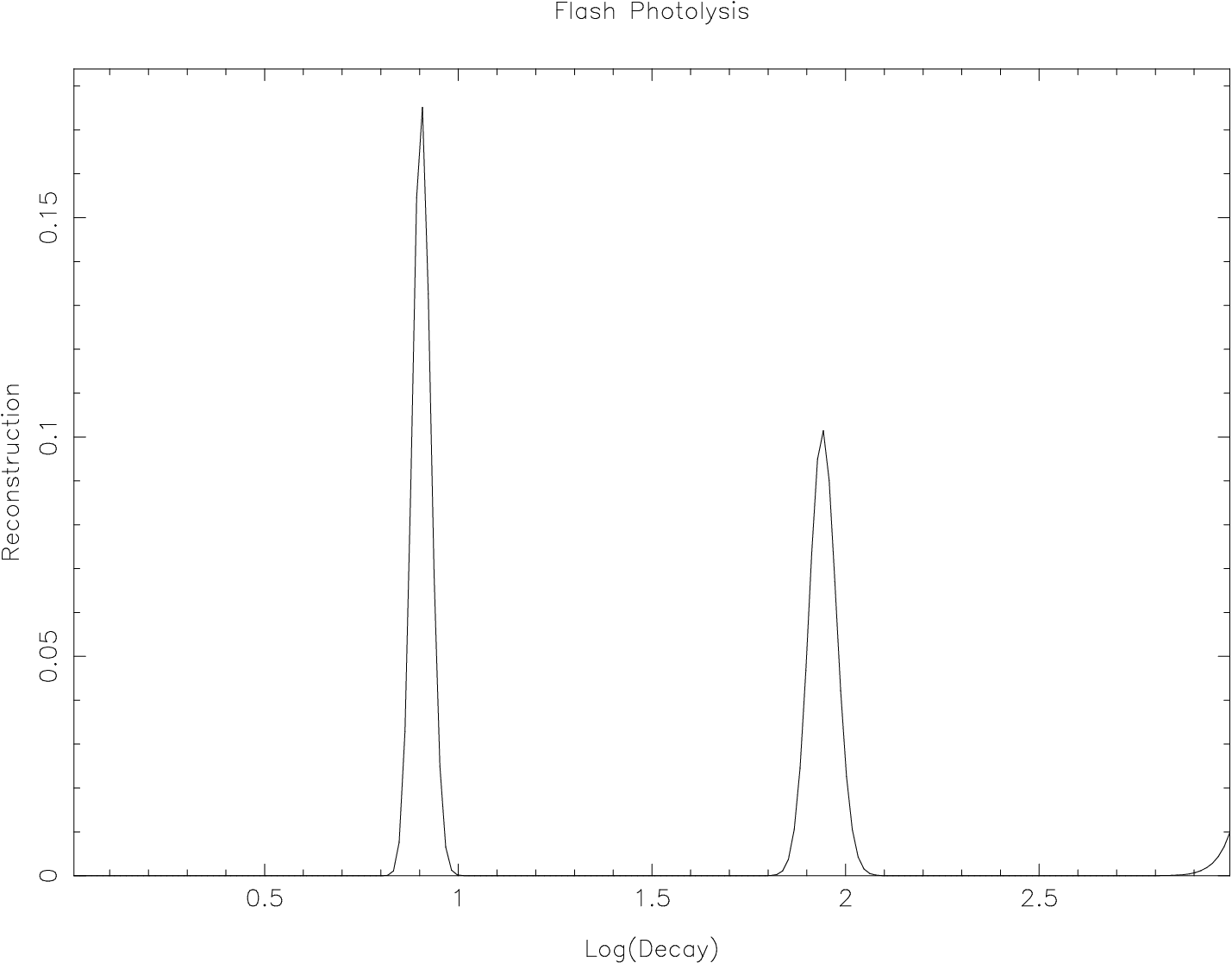}
\end{center}
\end{figure}

\begin{figure}
\begin{center}
\caption{The probability distribution for the number of distinct
decaying exponential components, as determined by MassInf}
\label{flash3}
\includegraphics[width=4cm,angle=-90]{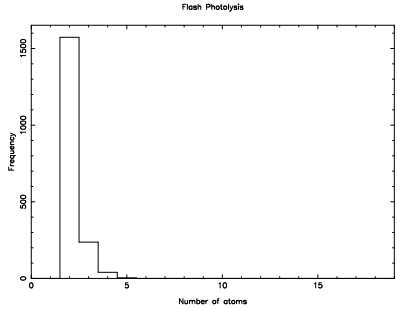}
\end{center}
\end{figure}

So far, some spectacular results have been
obtained for 1-dimensional spectra and 2-dimensional point sources.
An early application of Massive Inference has been to flash
photolysis data for proteins in corn grains, including a comparison
with MEM. Figure~\ref{flash1} shows simulations of the decay 
of luminescence measured in the experiment. The important question is how many
decaying exponentials are present in these data, and what are their
decay rates.
In Figure~\ref{flash2} the results obtained using MassInf and MEM are given.
We see that MassInf is far more successful in determining a
narrow range of possible decay rates. Most importantly, the MassInf algorithm
can
also provide the probability distribution for the number of distinct
exponential components, as shown in Figure~\ref{flash3}.

Table~\ref{comparison} summarises the similarities and differences
between MEM and Massive Inference.
\begin{table}
\centering
\caption{Comparison between of  MEM and Massive Inference}
\renewcommand{\arraystretch}{1.4}
\setlength\tabcolsep{5pt}
\begin{tabular}{cc}
\hline\noalign{\smallskip}
MEM & MassInf \\
\hline
Pixel based & Continuum \\
Gradient Search & Markov Chain Monte Carlo \\
Needs differential $\underline{R}(a)$ & Transform $R(f)$ only \\
and adjoint $\overline{R}(a)$ & \\
Poisson errors OK & $\chi^2$ only \\
Multi-dimensional & One-dimensional \\
 & (needs Peano curve) \\
Gaussian approximation & Direct sampling \\
Fast global transforms OK & Slow atom transforms \\
Computing time $\propto$ grid size & Computing time $\propto$ number atoms \\
\hline
\end{tabular}
\label{comparison}
\end{table}

\section{Conclusions}
\label{conc}

We have seen that a Bayesian approach provides a common statistical
framework for several important methods currently used in astronomical
processing and analysis. By generalising one's view to include also
the optimal choice of basis functions it is clear how significant
improvements can be obtained both in the quality of the resulting
reconstructions and in the speed at which the analysis can be carried
out. These two aspects will both be crucial in the new era of
quantitative cosmology which is now opening up.

\leftline{\bf Acknowledgments} We would like to thank 
Vlad Stolyarov, Sarah Bridle, Phil Marshall and Charlie McLachlan
for help with several
of the figures and Steve Gull for the simulations shown in 
Figures~\ref{flash1}--\ref{flash3}.

\clearpage
\addcontentsline{toc}{section}{Index}
\flushbottom
\printindex

\end{document}